\documentclass[prl,superscriptaddress,floatfix,letterpaper,twocolumn,showpacs]{revtex4-1}

\usepackage[utf8]{inputenc}

\newcommand{\ket}[1]{|#1\rangle}

\usepackage{amsmath}
\usepackage{amssymb}

\usepackage{graphicx}

\usepackage[urlcolor=blue, hyperindex, colorlinks, bookmarks=true]{hyperref}

\begin{document}

\title{Tunable coupling in circuit quantum electrodynamics \\ with a superconducting {V}-system}

\author{S.J. Srinivasan*}
\author{A.J. Hoffman*}
\affiliation{Department of Electrical Engineering, Princeton University, Princeton, NJ 08544, USA}
\author{J.M. Gambetta}
\affiliation{Institute for Quantum Computing and Department of Applied Mathematics, University of Waterloo, Waterloo, Ontario N2L 3G1, Canada}
\author{A.A. Houck}
\affiliation{Department of Electrical Engineering, Princeton University, Princeton, NJ 08544, USA}
\date{18 November, 2010.\\
 \**These authors contributed equally to this work}

\ifpdf
\DeclareGraphicsExtensions{.pdf, .jpg, .tif}
\else
\DeclareGraphicsExtensions{.eps, .jpg}
\fi

\maketitle

\textbf{Recent progress in superconducting qubits has demonstrated the potential of these devices for the future of quantum information processing.  One desirable feature for quantum computing is independent control of qubit interactions as well as qubit energies. We demonstrate a new type of superconducting charge qubit that has a V-shaped energy spectrum and uses quantum interference to provide independent control over the qubit energy and dipole coupling to a superconducting cavity. We demonstrate dynamic access to the strong coupling regime by tuning the coupling strength from less than 200 kHz to more than 40 MHz. This tunable coupling can be used to protect the qubit from cavity-induced relaxation and avoid unwanted qubit-qubit interactions in a multi-qubit system.}

The electromagnetic coupling between a quantized cavity field and a quantum mechanical two-level system has enabled the understanding of some of the most fundamental interactions between light and matter~\cite{Haroche2006}. Recent progress in this area has been spurred by the advent of circuit quantum electrodynamics (cQED), in which a superconducting qubit is strongly coupled to an on-chip microwave cavity~\cite{Wallraff2004}. This architecture can form the backbone of a superconducting quantum computer~\cite{Majer2007,Sillanpaa2007,Blais2007} and has been used to demonstrate efficient readout~\cite{Johnson,Vijay,Reed,Mallet2009}, complex entangled state preparation~\cite{DiCarlo,Neeley,Ansmann2009,Hofheinz2009}, and even elementary multi-qubit quantum processors~\cite{Neeley,DiCarlo2009,DiCarlo}. However, the same qubit-cavity coupling that enables efficient control and readout also reduces the qubit lifetime~\cite{Houck2008}, and also leads to spurious qubit-qubit crosstalk in a multi-qubit system. To avoid these detrimental effects, it is highly desirable to have dynamic control over the qubit-cavity coupling. Dynamic control has been demonstrated using an external coupling element between two directly coupled phase and flux qubits~\cite{Bialczak,Harris2007,Niskanen2007,Ploeg2007,Hime2006,Liu2006,Blais2003} and between a phase qubit and a lumped element resonator~\cite{Allman2009}. However, controllable coupling has so far eluded charge qubits such as the transmon because they cannot be coupled using magnetic flux, which is the mechanism employed by the external coupling elements in previous works. We develop a new charge qubit that uses quantum interference to provide an intrinsic method to control the coupling to a coplanar waveguide cavity.

This new variety of superconducting charge-based qubit, called the tunable coupling qubit (TCQ), has independent control over the qubit energy and the qubit-cavity coupling strength. It is based on a modified transmon design, and hence retains the essential charge noise insensitivity~\cite{Gambetta}. The quantum interference used to tune the coupling is controlled by applying two small magnetic fluxes with on-chip, fast flux bias lines~\cite{DiCarlo2009}. In this paper, we start by numerically solving the TCQ Hamiltonian, theoretically modelling the independent tunability of the qubit energy and coupling strength, $g$. Next, we show that cavity transmission measurements demonstrate a high degree of control over the qubit energy levels and coupling strengths. Finally, measurements of the vacuum Rabi splitting for several coupling strengths show that $g$ can be tuned from 45 MHz to values that are too small to be detected in this sample. This approach of using quantum interference to produce tunable coupling in a V-system should be readily generalizable to other finite anharmonic qubit systems.

\begin{figure}[!h]
	\includegraphics[scale=1.1]{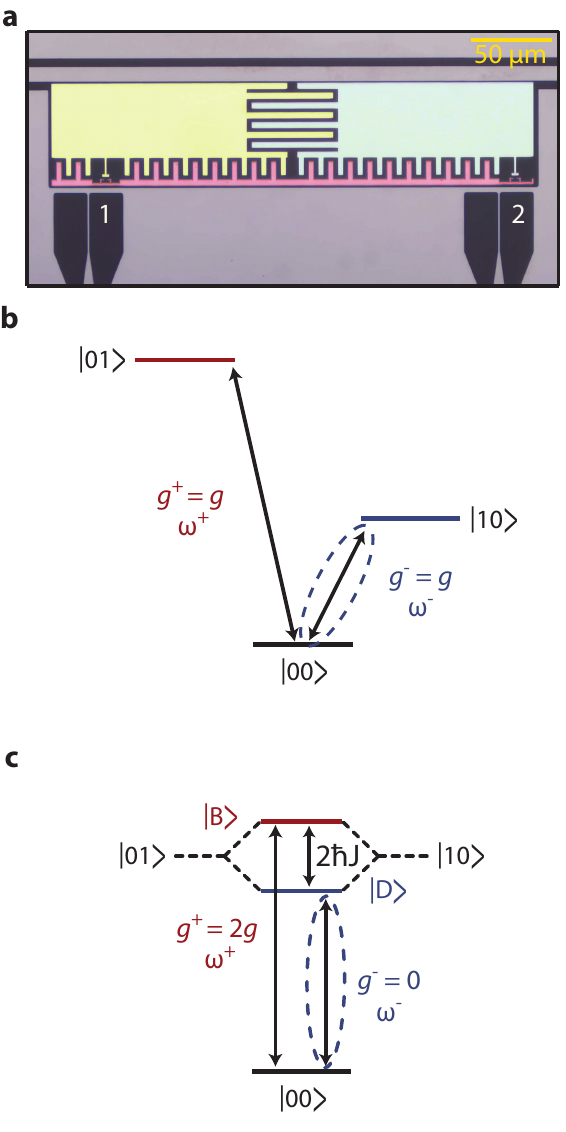}
	\caption{\label{DeviceImage} \textbf{Device image and level diagram.} \textbf{a}, False-colored, optical image of the TCQ coupled to a coplanar waveguide cavity.  The two flux bias lines, labeled ``1'' and ``2'', provide the field to change the Josephson energy of each split junction pair as $E_j = E_{j}^{\mathrm{max}} |\cos(\pi \phi/\phi_o)|$, where $E_{j}^{\mathrm{max}}$ is the maximum Josephson energy and is nominally equal for both split junction pairs. \textbf{b}, Energy level schematic for the situation when $|\omega^- - \omega^+| \gg J$.  Both qubit transitions behave independently and couple to the cavity with near-equal strength $g$.  \textbf{c}, Energy level schematic for when $\ket{01}$ and $\ket{10}$ are resonant.  Here, the energy levels hybridize to a bright state $\ket{B} = \frac{\ket{10} + \ket{01}}{\sqrt{2}}$ (red) and a dark state $\ket{D} = \frac{\ket{10} - \ket{01}}{\sqrt{2}}$ (blue) which couple to the cavity with coupling strengths $g^+ = 2g$ and $g^- = 0$, respectively.}
\end{figure}

The TCQ, shown in Fig. \ref{DeviceImage}a,
provides this dynamically tunable coupling and qubit energy. It consists of three coupled superconducting islands and two capacitively shunted SQUID loops, resulting in a V-shaped energy level spectrum shown in Fig. \ref{DeviceImage}b and c.  It can be understood as being composed of two independently controlled transmon qubits that are directly coupled to each other through a large capacitor, which determines the characteristic interaction energy, $\hbar J$.  The resulting spectrum has two collective excitations with transition frequencies $\omega^-$ and $\omega^+$, and coupling strengths, $g^-$ and $g^+$, respectively.  When $|\omega^+ - \omega^-| \gg J$, as in Fig. \ref{DeviceImage}b, the two excited states are essentially the independent transmon eigenstates $\ket{01}$ and $\ket{10}$, and the coupling strengths are determined primarily by the qubit geometry. In this regime, for the device presented here, $g^- = g^+ = g$; however, in general, $g^+$ and $g^-$ need not be identical and are determined by the capacitances of each island to the center pin of the cavity. When the two excited energy levels are near-degenerate, they hybridize to give a superposition of the independent energy levels as shown in Fig. \ref{DeviceImage}c.  At resonance, the energy level splitting is $\hbar J$ and the excited states of the qubit are $|B\rangle = \frac{|01\rangle + |10\rangle}{\sqrt{2}}$ and $|D\rangle = \frac{|01\rangle - |10\rangle}{\sqrt{2}}$.  These two states couple to the cavity very differently.  The bright state, $\ket{B}$, has a strong dipole moment with an enhanced coupling, $2g$. On the other hand, the dark state, $\ket{D}$ has a quadrupole moment and has zero dipole coupling.

This device can be used as a qubit by restricting operation to the two lowest energy levels, which is denoted in this paper by the $\omega^-$ transition. The ground state is $\ket{00}$, while the excited state varies continuously from $\ket{10}$ ($g^- = g$) to $\ket{D}$ ($g^- = 0$) depending on the desired dipole coupling for the transition. The other excited state is used only to generate quantum interference by bringing it into resonance with the first excited state, canceling its dipole coupling.  Using the two independent flux controls, it is possible to move adiabatically between the resonant and non-resonant cases and thus tune the qubit coupling. Because flux is applied with fast flux bias lines, $\omega^-$ and $g^-$ can be tuned rapidly.

\begin{figure}[h]
	\includegraphics[scale=1.4]{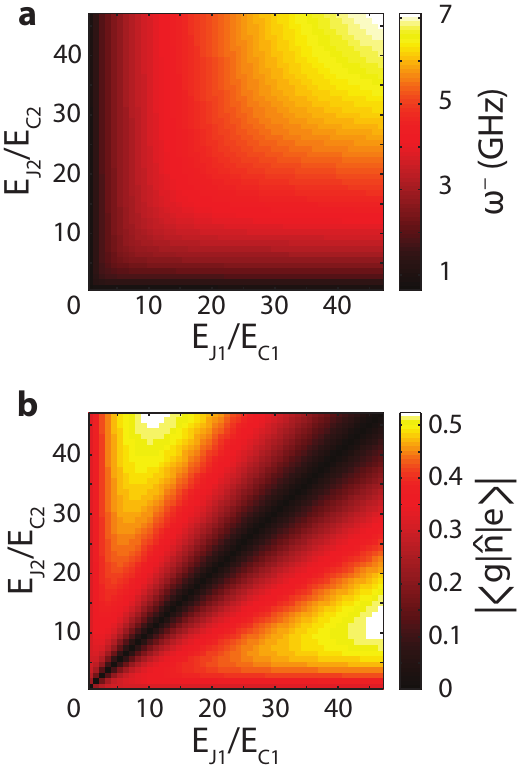}
	\caption{\label{Calculations} \textbf{Simulated qubit parameters.} \textbf{a},  Calculated transition frequency between the two lowest energy states using equation (\ref{eq:Hamiltonian}) with $E_i = 369\, \mathrm{MHz}$, $E_{c1} = 428 \, \mathrm{MHz}$, and $E_{c2} = 429 \, \mathrm{MHz}$ for a range of $E_{J1}$ and $E_{J2}$.  When $E_{J1}$ and $E_{J2}$ are significantly different, the smaller Josephson energy is the primary factor in determining the qubit transition.  \textbf{b}, Calculated dipole coupling matrix element, $|\langle g|\hat{n}|e \rangle|$ versus $E_{J1}$ and $E_{J2}$.  The qubit-cavity coupling for the $\omega^-$ transition, $g^-$, is directly proportional to the plotted matrix element. For the qubit transition, $\omega^-$, coupling between the qubit and cavity is suppressed along the diagonal when the $\ket{01}$ and $\ket{10}$ states are resonant.}
\end{figure}

The TCQ has been explored in greater theoretical detail in reference~\cite{Gambetta}. Here we study the quantitative behavior by numerically solving the Hamiltonian:
\begin{gather} \label{eq:Hamiltonian}
H = 4 E_{c1} (n_1 - n_{g1})^2 + E_{J1} \cos(\gamma_1) + 4 E_{c2} (n_2 - n_{g2})^2 \notag \\
+ E_{J2} \cos(\gamma_2) + 4 E_i n_1 n_2
\end{gather}
The first four terms describe two independent anharmonic energy ladders and the last term represents the interaction between these ladders due to the capacitive coupling of the top islands. $E_{J1}$ and $E_{J2}$ are the Josephson energies for each of the two SQUID loops and $\gamma_1$ and $\gamma_2$ are the superconducting phases across the junctions; $E_{c1}$ and $E_{c2}$ are charging energies for the two islands; $ n_1$ ($n_2$) and $n_{g1}$ ($n_{g2}$) are the number of cooper pairs and the effective charge of an applied gate voltage, respectively, on the left (right) island; $E_i$ is called the interaction energy and its magnitude determines the resonant splitting,  $2 \hbar J = \sqrt{2}E_i(\frac{E_{J1}}{E_{c1}})^{1/4}(\frac{E_{J2}}{E_{c2}})^{1/4}$, between the $\omega^-$ and $\omega^+$ transitions. Numerical diagonalization of the Hamiltonian is used to produce plots of the qubit transition energy and the dipole matrix coupling element shown in Fig. \ref{Calculations}.
The qubit's dipole coupling for the symmetric design presented here can be calculated as $\hbar g^- = 2 e^2 \beta V^0_{rms} |\langle g|\hat{n}|e \rangle|$, where $\ket{g}$ is the ground state, $\ket{e}$ is the first excited state, $\hat{n} = \hat{n}_1 + \hat{n}_2$ is the charge operator, $V^0_{rms}$ is the root mean square of the voltage fluctuations in the cavity, and $\beta$ is the qubit capacitance division~\cite{Gambetta}. The plots show that $g^-$ and $\omega^-$ tune in nearly-orthogonal directions with changes in the Josephson energies of the two junctions.  This relationship between $g^-$ and $\omega^-$ allows these two device parameters to be \emph{independently} controlled by application of appropriate flux to the SQUID loops.

To demonstrate this independent tunability, a TCQ is coupled to a superconducting coplanar waveguide cavity which is fabricated using conventional processing techniques on a 200 nm thick Nb film sputtered on a sapphire substrate.  The cavity used here has a bare $\lambda/2$ resonance at $\omega_c = 5.784$ GHz and a full-width at half maximum of $\kappa = 1.2$ MHz. The three island device is fabricated in a notch situated near the anti-node of the cavity using electron beam lithography and a double-angle Al evaporation~\cite{Frunzio2005}. The device is cooled to 25 mK in a dilution refrigerator. A pair of low-noise, current-biased voltage sources are used to independently control $E_{J1}$ and $E_{J2}$ via two separate flux bias lines.  A two-port network analyzer is used to measure transmission, $S_{21}$, at few-photon power levels versus frequency over a range of independent flux bias voltages.

\begin{figure}[h]
	\includegraphics[scale=1.2]{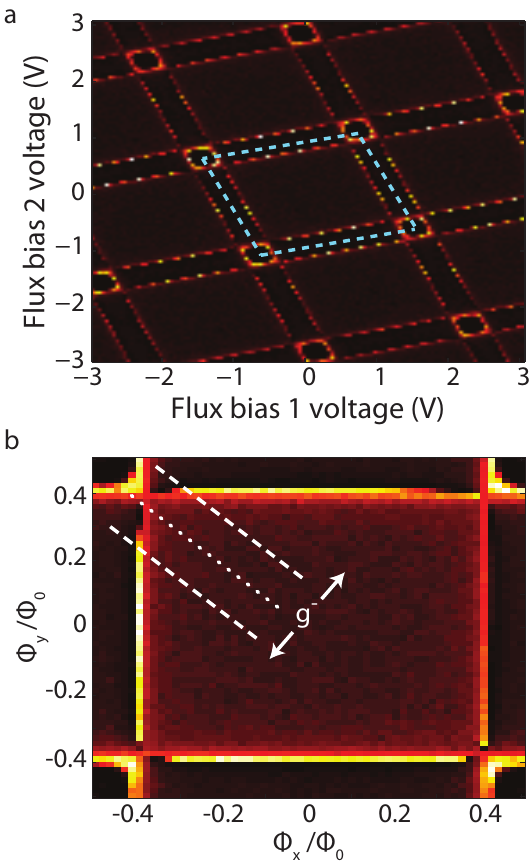}
	\caption{\label{Slices} \textbf{Cavity transmission.} \textbf{a}, Transmission at 5.777 GHz is shown over several flux quanta. The two directions of periodicity are nonorthogonal due to a finite current coupling between the flux bias lines. The blue parallelogram indicates a single flux quantum. \textbf{b}, The axes are orthogonalized via a linear transformation on the control voltage and transmission is measured at 5.777 GHz over a single flux quantum. The bright horizontal and vertical lines of high transmission at 5.777 GHz are due to a shift in the cavity from its interaction with the $\ket{01}$ and $\ket{10}$ states. The center of the plot indicates the point of maximum $E_{J1}$ and $E_{J2}$, and therefore maximum $\omega^-$. The overlaid white diagonal lines indicate contours of constant $g^-$. The dotted line shows the contour of minimum $g^-$ and the white arrows point in the directions of contours with higher $g^-$.}
\end{figure}

The TCQ is characterized by varying both flux biases over several flux quanta and measuring the cavity transmission at 5.777 GHz, as shown in Fig. \ref{Slices}a. 
The energy levels of the coupled qubit-cavity system tune into and out of resonance with the applied microwave tone, causing the transmission to vary dramatically. A finite current coupling between the flux bias lines causes the two axes of periodicity in the plot to be non-orthogonal. When this skew is corrected and the flux controls are swept over a single period, cavity transmission at 5.777 GHz shows prominent horizontal and vertical lines, indicating points where the $\ket{01}$  or $\ket{10}$ state is near resonant with the cavity. When the two transitions are resonant with each other, indicated by the white dotted line connecting the intersection points of the horizontal and vertical lines, the $\ket{B}$ and $\ket{D}$ states are the energy eigenstates of the qubit. Along the white dotted line, the qubit energy $\omega^-$ varies, but $g^-$ is zero. The coupling increases symmetrically in an orthogonal direction about the dotted line, as indicated by the two arrows pointing to the dashed contours.

\begin{figure}[h]
	\includegraphics[scale=.65]{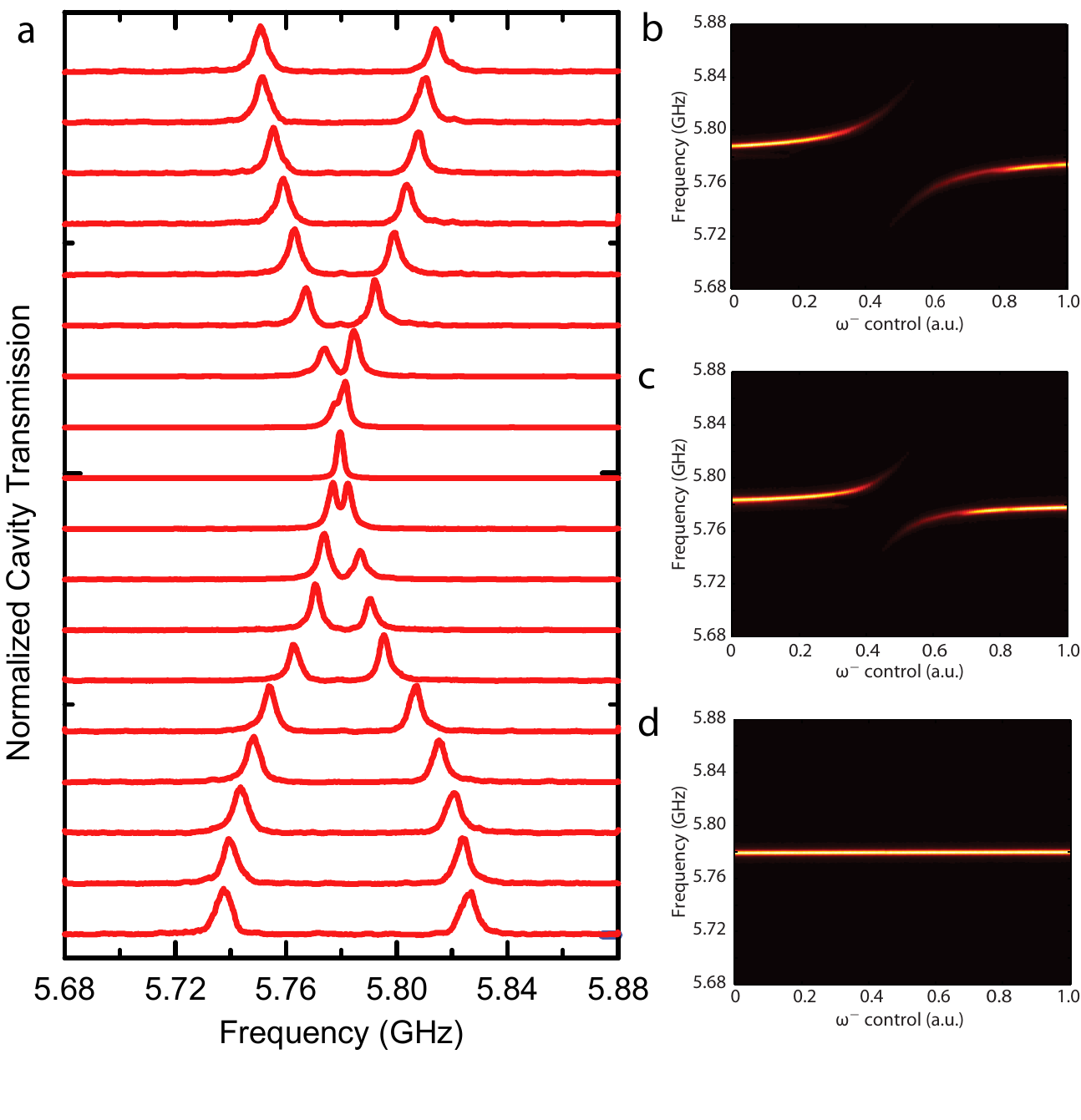}
	\caption{\label{VRS} \textbf{Tunable qubit-cavity coupling.} \textbf(a), Normalized transmitted power at resonance for $g$ tuning from 45 MHz, through the $g=0$ point, to 33 MHz. (b-d) show the $\omega^-$ transition tuning through the cavity resonance for $g = 45 $MHz, $g = 27 $MHz, and $g < 200 $kHz, respectively. The cavity linewidth places an upper bound on the resolution with which one can observe the minimum splitting on resonance.}
\end{figure}

The transmission measurements in Fig. \ref{VRS}a, where the qubit is resonant with the cavity ($\omega^- = \omega_c$), clearly demonstrate tunable coupling.
Several vacuum Rabi crossings are shown with controllable splitting.  In the waterfall data, the coupling strength is reduced from a maximum value of 43 MHz to an imperceptibly small value, less than 200kHz, with resolution limited by the finite cavity linewdith.  After crossing through zero, the coupling increases again to larger values.  Fig. \ref{VRS}b-d show transmission while tuning the qubit frequency through resonance while holding $g^-$ constant for three different values of $g^-$.  For the large and mid-range $g^-$, avoided crossings are observed as expected.  In Fig. \ref{VRS}d, no change in the cavity line is observed as the qubit is tuned through resonance when the coupling is zero.  From this data, it is readily apparent that the device operates as expected, providing independently tunable frequency and coupling with on chip fast flux bias lines.

Tunable coupling alleviates two major challenges in cQED-based experiments: radiative qubit decay and qubit-qubit crosstalk. As a result of the interaction between the cavity and qubit, the qubit states are dressed with a photonic component proportional to $\frac{g}{\Delta}$, where the detuning, $\Delta = \omega_{q} - \omega_{c}$, is the difference between the bare qubit and cavity frequencies. The dressed qubit state has a direct radiative pathway for relaxation through the cavity, known as the Purcell effect. This has been shown to be a major factor limiting lifetimes of transmon qubits~\cite{Houck2008}. Careful engineering of the admittance of the superconducting circuit can suppress relaxation~\cite{Reed2010}, but may prove difficult in large scale circuits and does not help with qubit-qubit crosstalk. Both crosstalk and relaxation can be suppressed by operating at large detunings~\cite{DiCarlo2009}, but this becomes increasingly difficult with a large number of qubits and leaves a finite coupling that still causes errors. Moreover, small detuning is often required for high-fidelity measurements. The TCQ's tunable coupling is a viable solution to both of these problems because the coupling can be turned off when spurious interactions need to be suppressed. Additionally, the TCQ is a V-system and has a second excited state, represented here by the $\omega^+$ transition. This transition remains at least 1 GHz detuned from the primary qubit transition and could prove beneficial in a high SNR cycling measurement of the qubit state~\cite{Gambetta,Monroe1995,Yang1989}. Considering the unique energy level configuration and the benefits that dynamically tunable coupling provides, the TCQ is an important step in enabling large-scale quantum processors.

\begin{acknowledgements}
	The authors acknowledge useful discussions with David I. Schuster, Lev S. Bishop, Alexandre Blais, and Robert J. Schoelkopf. The Princeton work is supported by the Alfred P. Sloan Foundation, the David and Lucile Packard Foundation, and by DARPA (award number HR0011-06-D-0003). IQC work is supported by NSERC and CIFAR.  SJS, AJH, and AAH designed the experiment, fabricated samples, acquired data, analyzed results, and prepared the manuscript. JMG provided theory support and aided in device design and analysis.
\end{acknowledgements}

\bibliographystyle{ieeetr}
\bibliography{tunablecoupling}
\end{document}